\global\def\draftcontrol{0}

   \def\versionno{ alpha prime mistake -- draft   }

\catcode`\@=11

\expandafter\ifx\csname draftcontrol\endcsname\relax\global\def\draftcontrol{0}
\fi

{\count255=\time\divide\count255 by 60
\xdef\hourmin{\number\count255}
\multiply\count255 by-60\advance\count255 by\time
\xdef\hourmin{\hourmin:\ifnum\count255<10 0\fi\the\count255}}
\def\draftdate{\number\month/\number\day/\number\year\ \ \ \hourmin }

\newcommand\makepapertitle{\par
  \begingroup
    \renewcommand\thefootnote{\@fnsymbol\c@footnote}%
    \def\@makefnmark{\rlap{\@textsuperscript{\normalfont\@thefnmark}}}%
    \long\def\@makefntext##1{\parindent 1em\noindent
            \hb@xt@1.8em{%
                \hss\@textsuperscript{\normalfont\@thefnmark}}##1}%
     \newpage
     \global\@topnum\z@   
     \@makepapertitle
     \thispagestyle{empty}\@thanks
  \endgroup
  \setcounter{footnote}{0}%
  \global\let\thanks\relax
  \global\let\makepapertitle\relax
  \global\let\@makepapertitle\relax
  \global\let\@thanks\@empty
  \global\let\@author\@empty
  \global\let\@date\@empty
  \global\let\@title\@empty
  \global\let\title\relax
  \global\let\author\relax
  \global\let\date\relax
  \global\let\and\relax
  \def\version{\let\version\@version\@gobble}
}
\def\@makepapertitle{%
  \newpage
   \ifnum\draftcontrol=1 {}
   \version\versionno
   \vskip 3em%
   \else
   \hfill\hbox to 3cm {\parbox{4cm}{\@pubnum}\hss}%
   \vskip 3em%
   \fi
   \begin{center}%
   \let \footnote \thanks
     {\LARGE {\@title}}%
     \vskip 1.5em%
     {\normalsize
       \lineskip .5em%
       \begin{tabular}[t]{c}%
         \@author
       \end{tabular}\par}%
     \vskip 1.5em%
     {\@bstract}%
     \end{center}%
     \vskip 1.5em
     \@date%
   \par
}

\gdef\@pubnum{}
\def\pubnum#1{%
  \gdef\@pubnum{#1}}

\gdef\@bstract{}
\def\Abstract#1{%
  \gdef\@bstract{%
   \parbox{\textwidth-0pc}{%
   \centerline{\bf Abstract}\penalty1000%
\kern.2cm%
\noindent
\renewcommand\baselinestretch{1.0}%
{#1}}}
}

\def\ps@paper{\let\@mkboth\@gobbletwo%
     \ifnum\draftcontrol=1
	\def\@oddfoot{\hbox to \textwidth{\tiny \versionno \hfil\tiny\draftdate}%
	\hskip -\textwidth \hbox to \textwidth{\hfil\rm\thepage\hfil}}%
     \else\def\@oddfoot{\hbox to \textwidth{\hfil\rm\thepage\hfil}}
     \fi
     \let\@evenfoot\@oddfoot
}

\def\body{\clearpage
          \pagestyle{paper}
	}

\def\@version#1{\ifnum\draftcontrol=1
\typeout{}\typeout{#1}\typeout{}
\vskip3mm\centerline{\hbox{\fbox{\normalsize{\tt DRAFT -- #1 -- }
                   {\draftdate}}}}\vskip3mm
\fi}
\let\version\@version
\long\def\eqlabel#1{\ifnum\draftcontrol=1
                    \tag@false  
                    \tag*{(\theequation) \hbox to -0.2cm{\hspace{0cm}\small{#1}\hss}}
                    \refstepcounter{equation}
                    \edef\@currentlabel{\theequation}
                    \ltx@label{#1}          
                    \else
                    \label{#1}
                    \fi
                    }
\let\st@bibitem\@bibitem
\let\st@lbibitem\@lbibitem
\ifnum\draftcontrol=1
  \def\@bibitem#1{%
    \st@bibitem{#1}\a@@label{#1}\ignorespaces}
  \def\@lbibitem[#1]#2{%
    \st@lbibitem[#1]{#2}\a@@label{#2}\ignorespaces}
  \def\a@@label#1{%
    \gdef\a@lab{\smash{\normalfont\small#1}}
    \ifvmode
      \if@inlabel
        \global\setbox\@labels\hbox{%
          \llap{\a@lab\let\a@lab\relax
                \kern\@totalleftmargin\kern\marginparsep}%
          \box\@labels}%
      \fi
    \fi}
\fi

\documentclass[12pt,letterpaper]{article}

\usepackage{amsmath,amssymb,array,calc,rotating,epsfig,psfrag}
\usepackage[nosort]{cite}

\ifnum\draftcontrol=1
\fi

\renewcommand\baselinestretch{1.25}
\renewcommand\section{\@startsection {section}{1}{\z@}%
                                   {-3.5ex \@plus -1ex \@minus -.2ex}%
                                   {2.3ex \@plus.2ex}%
                                   {\normalfont\large\bfseries}}
\renewcommand\subsection{\@startsection{subsection}{2}{\z@}%
                                   {-3.25ex\@plus -1ex \@minus -.2ex}%
                                   {1.5ex \@plus .2ex}%
                                   {\normalfont\normalsize\bfseries}}
\renewcommand\subsubsection{\@startsection{subsubsection}{3}{\z@}%
                                   {-3.25ex\@plus -1ex \@minus -.2ex}%
                                   {1.5ex \@plus .2ex}%
                                   {\normalfont\normalsize\it}}
\renewcommand\paragraph{\@startsection{paragraph}{4}{\z@}%
                                   {-3.25ex\@plus -1ex \@minus -.2ex}%
                                   {1.5ex \@plus .2ex}%
                                   {\normalfont\normalsize\bf}}

\numberwithin{equation}{section}

\def\ie{{\it i.e.}}

\def\revise#1       {\raisebox{-0em}{\rule{3pt}{1em}}%
                     \marginpar{\raisebox{.5em}{\vrule width3pt\
                     \vrule width0pt height 0pt depth0.5em
                     \hbox to 0cm{\hspace{0cm}{%
                     \parbox[t]{4em}{\raggedright\footnotesize{#1}}}\hss}}}}

\def\calc         {{\cal C}}

\def\caln         {{\cal N}}
\def\calo         {{\cal O}}

\def\sqr#1#2{{\vcenter{\vbox{\hrule height.#2pt
 \hbox{\vrule width.#2pt height#1pt \kern#1pt
 \vrule width.#2pt}\hrule height.#2pt}}}}

\def\a{\alpha}
\def\w{\omega}
\def\g{\gamma}

\newcommand{\qq}{\mathfrak{q}}
\newcommand{\ww}{\mathfrak{w}}

\def\l{\lambda}

\catcode`\@=12

\begin{document}


\title{Resolving disagreement for $\eta/s$ in a CFT plasma at finite coupling}

\pubnum{%
UWO-TH-08/8
}
\date{May 2008}

\author{
Alex Buchel\\[0.4cm]
\it Department of Applied Mathematics\\
\it University of Western Ontario\\
\it London, Ontario N6A 5B7, Canada\\[0.2cm]
\it Perimeter Institute for Theoretical Physics\\
\it Waterloo, Ontario N2J 2W9, Canada\\
}

\Abstract{
The ratio of shear viscosity to entropy density in a strongly coupled
CFT plasma can be computed using the AdS/CFT correspondence either
from equilibrium correlation functions or from the Janik-Peschanski
dual of the boost invariant plasma expansion.  We point out that the
previously found disagreement for $\eta/s$ at finite t' Hooft coupling
is resolved once the incoming-wave boundary condition for metric
fluctuations at the horizon of the dual geometry is properly imposed.
}


\makepapertitle

\body

\version\versionno

\section{Introduction}
Gauge theory/string theory correspondence of Maldacena \cite{m9711,m2} has been useful in analysis of the transport properties 
of the strongly coupled gauge theory plasma \cite{ss}. In particular, it was proven that the ratio of shear viscosity to the 
entropy density $\frac{\eta}{s}$ at infinite 't Hooft coupling is universal in all gauge theory plasmas which allow 
for a holographically dual  
string theory description \cite{u1,u2,u3,u4}. At finite t' Hooft coupling (but still in the planar limit), 
this ratio receives leading contribution from $\calo(\a'^3)$ string theory corrections to the dual type IIB supergravity 
background. In \cite{ab} it was argued that such corrections are universal as well, as long as the dual gauge theory plasma 
is conformal. 

The correction to the ratio $\frac{\eta}{s}$ can be computed either from equilibrium correlation functions 
(as in \cite{f1,f2}) or by imposing a non-singularity condition of the $\calo(\a'^3)$ corrected 
Janik-Peschanski \cite{jp} dual of the boost invariant CFT plasma expansion (as in \cite{f3}).
In the former case it was found that \cite{f1,f2}
\begin{equation}
\frac{\eta}{s}=\frac{1}{4\pi}\left(1+\frac{135}{8}\ \zeta(3)\ \l^{-3/2}+\cdots\right)\,,
\eqlabel{o1}
\end{equation}
while in the latter \cite{f3}
\begin{equation}
\frac{\eta}{s}=\frac{1}{4\pi}\left(1+\frac{120}{8}\ \zeta(3)\ \l^{-3/2}+\cdots\right)\,,
\eqlabel{o2}
\end{equation}
where $\l$ is the $\caln=4$ supersymmetric Yang-Mills 't Hooft coupling.

In this paper we resolve the discrepancy between \eqref{o1} and \eqref{o2}. It turns out  
that the incoming-wave boundary condition on metric fluctuations used to obtain \eqref{o1}
were imposed at the supergravity level, rather than at $\calo(\a'^3)$ string theory corrected background.
In what follows we show that once the boundary conditions are properly imposed, the shear viscosity to the 
entropy ratio obtained from equilibrium correlation functions agrees with \eqref{o2}.

\section{Incoming wave boundary condition}
We consider here the shear quasinormal mode in $\calo(\a'^3)$ near-extremal D3 brane geometry. Discussion extends to both 
the sound quasinormal mode and the scalar quasinormal mode. 

Equations of motion to the shear quasinormal mode in  $\calo(\a'^3)$ near-extremal D3 brane geometry were derived in \cite{f2}. 
These equations can be expanded perturbatively 
in $\g\equiv \frac 18 \zeta(3)\ (\a')^3$, provided we introduce
\begin{equation}
Z_{shear}=Z_{shear,0}+\g\ Z_{shear,1}+\calo(\g^2)\,.
\eqlabel{zshear}
\end{equation}
We find
\begin{equation}
\begin{split}
0=&Z_{shear,0}''+\frac{x^2\qq^2+\ww^2}{x(\ww^2-x^2 \qq^2)}\ Z_{shear,0}'+\frac{\ww^2-x^2 \qq^2}{x^2(1-x^2)^{3/2}}\ 
Z_{shear,0}\,,\\
0=&Z_{shear,1}''+\frac{x^2\qq^2+\ww^2}{x(\ww^2-x^2 \qq^2)}\ Z_{shear,1}'+\frac{\ww^2-x^2 \qq^2}{x^2(1-x^2)^{3/2}}\ 
Z_{shear,1}+J_{shear,0}\left[Z_{shear,0}\right]\,,
\end{split}
\eqlabel{zsea1}
\end{equation}
where the source  $J_{shear,0}$ is a functional of the zero's order shear mode $Z_{shear,0}$
\begin{equation}
\begin{split}
J_{shear,0}\left[Z_{shear,0}\right]=&\calc_{shear}^{(4)}  \frac{d^4 Z_{shear,0}}{d x^4}+\calc_{shear}^{(3)}
\ \frac{d^3 Z_{shear,0}}{d x^3}+\calc_{shear}^{(2)}
\  \frac{d^2 Z_{shear,0}}{d x^2}\\
&+\calc_{shear}^{(1)}\ \frac{d Z_{shear,0}}{d x}+\calc_{shear}^{(0)}\ Z_{shear,0}\,.
\end{split}
\eqlabel{sourceshear}
\end{equation}
The coefficients $\calc_{shear}^{(i)}$ are given explicitly in appendix A of \cite{f2}.
In \eqref{zsea1} we introduced 
\begin{equation}
\ww=\frac{\w}{2\pi T_0}\,,\qquad \qq=\frac{q}{2\pi T_0}\,.
\eqlabel{wwqqdef}
\end{equation}
where $T_0$ is a near-extremal D3 brane temperature in the supergravity approximation. 

At the supergravity level, \ie, for $Z_{shear,0}$, the incoming-wave boundary condition at the horizon implies that 
in the hydrodynamic approximation  
\begin{equation}
\begin{split}
Z_{shear,0}=&x^{-i\ww}\ \left(z_{shear,0}^{(0)}+i\qq z_{shear,0}^{(1)}+\calo(\qq^2)\right)\,,\\
\end{split}
\eqlabel{sss1a}
\end{equation} 
where $z_{shear,0}^{(i)}$ are regular at the horizon. While it is possible to use ansatz 
\begin{equation}
\begin{split}
Z_{shear,1}=&x^{-i\ww}\ \left(z_{shear,1}^{(0)}+i\qq z_{shear,1}^{(1)}+\calo(\qq^2)\right)\,,\\
\end{split}
\eqlabel{sss1b}
\end{equation} 
with regular $z_{shear,1}^{(i)}$ at the horizon ( as was done in \cite{f1,f2} ) to order $\calo(\qq)$, it is straightforward 
to verify that $\calo(\qq^2)$ term in \eqref{sss1b} is always singular.  The reason for this is that the asymptotic 
$\propto x^{-i\ww}$ is an incoming-wave boundary condition only at the supergravity level, but is modified at $\calo(\g)$. 

To determine the correct incoming-wave boundary condition we have to go back to the equation of motion for $Z_{shear}$ \eqref{zshear}:
\begin{equation}
\begin{split}
0=&Z_{shear}''+\frac{x^2\qq^2+\ww^2}{x(\ww^2-x^2 \qq^2)}\ Z_{shear}'+\frac{\ww^2-x^2 \qq^2}{x^2(1-x^2)^{3/2}}\ 
Z_{shear}+J_{shear,0}\left[Z_{shear}\right]+\calo(\g^2)\,.
\end{split}
\eqlabel{mod}
\end{equation}
We look for solution to \eqref{mod} to order $\calo(\g)$ within the ansatz 
\begin{equation}
\begin{split}
Z_{shear}=&x^{\beta}\ \left(z_{shear}^{(0)}+i\qq z_{shear}^{(1)}+\calo(\qq^2)\right)\times\left(1+\calo(\g^2)\right)\,,\\
\end{split}
\eqlabel{sss1c}
\end{equation} 
with regular  $z_{shear}^{(i)}$ at the horizon.
Substituting \eqref{sss1c} into \eqref{mod} we find (to order $\calo(\g)$):
\begin{equation}
0=x^{\beta}\left(\frac{1}{x^2}\ \left(\beta^2+\ww^2(1-30 \g)\right)+\calo(x^0)\right)\,.
\eqlabel{correction}
\end{equation}
From \eqref{correction} we see that the incoming wave boundary condition is 
\begin{equation}
\beta=-i\ww (1-15\g)+\calo(\g^2)\,.
\eqlabel{beta}
\end{equation}
Eq.~\eqref{beta} is  an expected modification. Indeed, computations of the spectrum of shear and sound quasinormal modes in
more complicated supergravity backgrounds \cite{com1,com2}, as well as the general arguments for the scalar quasinormal mode 
in \cite{u3}, show that the incoming-wave boundary condition at the horizon always takes the form $\propto x^{-i \frac{\omega}{2\pi T}}$. 
Thus, a $(1-15\g)$ rescaling of the supergravity boundary conditions is simply a well-known rescaling of the near-extremal 
D3 brane temperature due to string theory higher derivative corrections \cite{gkt1}:
\begin{equation}
T=T_0 (1+15\g+\calo(\g^2))\,.
\eqlabel{temp}
\end{equation}
 
We emphasize again that the boundary condition \eqref{beta} is required not only to obtain correct physical results, 
but it is  mandatory if one attempts to extend computation of the spectrum of quasinormal modes beyond the first order in 
the hydrodynamic approximation.

\section{Corrected shear and sound quasinormal modes to first order in the hydrodynamic approximation}

In previous section we argued that the incoming-wave boundary condition for the near-extremal D3 brane
quasinormal modes receives order $\calo(\a'^3)$ correction.  
It turns  out that  this  modification \eqref{beta} is the source of the discrepancy between \eqref{o1} and \eqref{o2}. 
To show the latter we have to recompute the spectrum of the shear and sound quasinormal modes. It is straightforward to do so 
following detailed discussion in \cite{f2}. 

For the shear quasinormal mode we find 
\begin{equation}
\begin{split}
z_{shear,0}^{(0)}=1\,,\qquad z_{shear,0}^{(1)}=\frac 12 \frac {\qq}{\ww} x^2\,,
\end{split}
\eqlabel{sss3}
\end{equation}
\begin{equation}
\begin{split}
z_{shear,1}^{(0)}=&\frac{25}{16}x^2\left(x^4-4x^2+5\right)\,,
\\
z_{shear,1}^{(1)}=&-\frac{1}{32\qq \ww}x^2\biggl(\qq^2\left(-240-1565 x^2-860 x^4+695x^6\right)\\
&+16\ww^2
\left(594-264x^2+43x^4\right)\biggr)+\delta z_{shear,1}^{(1)}\,,
\end{split}
\eqlabel{sss4}
\end{equation}
and for the sound quasinormal mode we find
\begin{equation}
\begin{split}
z_{sound,0}^{(0)}=\frac{3\ww^2+(x^2-2)\qq^2}{3\ww^2-2\qq^2}\,,\qquad z_{sound,0}^{(1)}=\frac {2\ww\qq x^2}
{3\ww^2-2\qq^2}\,,
\end{split}
\eqlabel{ssss3}
\end{equation}
\begin{equation}
\begin{split}
z_{sound,1}^{(0)}=&\frac{5x^2}{16(3\ww^2-2\qq^2)^2}\biggl(\qq^4\left(2404+446x^2-4164x^4+2006x^6\right)
\\
&-3\ww^2\qq^2\left(1588+183x^2-2072x^4+1003x^6\right)+45\ww^4\left(5-4x^2+x^4\right)
\biggr)\,,
\\
z_{sound,1}^{(1)}=&\frac{\ww x^2}{8\qq (3\ww^2-2\qq^2)^2}\biggl(\qq^4\left(-13344+5846x^2-4520x^4+1734x^6\right)
\\
&-3\ww^2\qq^2\left(-9744+5035x^2-2604x^4+867x^6\right)\\
&-36\ww^4\left(594-264x^2+43x^4\right)
\biggr)+\delta z_{sound,1}^{(1)}\,,
\end{split}
\eqlabel{ssss4}
\end{equation}
where $\delta z_{shear,1}^{(1)}$ and $\delta z_{sound,1}^{(1)}$ are corrections due to the modified boundary condition \eqref{beta}:
\begin{equation}
\delta z_{shear,1}^{(1)}=-\frac{15\qq x^2}{2\ww}\,,\qquad \delta z_{sound,1}^{(1)}=\frac{30\ww\qq x^2}{2\qq^2-3\ww^2}\,.
\eqlabel{deltas}
\end{equation}

Imposing the Dirichlet condition on $x^{i\ww(1-15\g)} Z_{shear,0} $ and  $x^{i\ww(1-15\g)} Z_{sound,0} $ at the boundary 
determines the lowest shear and sound quasinormal frequencies
\begin{equation}
\begin{split}
{\rm shear}:&\qquad \ww=-i\qq^2\left(\frac 12+\frac{105}{2}\g\right)+\calo(\qq^3,\g^2)\,, \\
{\rm sound}:&\qquad \ww=\frac{1}{\sqrt{3}}\qq-i\qq^2\left(\frac 13+\frac{105}{3}\g\right)+\calo(\qq^3,\g^2)\,. \\
\end{split}
\eqlabel{results}
\end{equation}
Note that both channels lead to the same prediction for $\frac{\eta}{s}$, namely, the one given by \eqref{o2}.
Additionally, as expected, neither the speed of sound nor the bulk viscosity receives $\calo(\g)$ corrections, 
which are forbidden by the conformal symmetry.

\section*{Acknowledgments}
I would like to thank Miguel Paulos for valuable discussions.
My research at Perimeter Institute is supported in part by the Government
of Canada through NSERC and by the Province of Ontario through MRI.
I gratefully acknowledges further support by an NSERC Discovery
grant and support through the Early Researcher Award program by the
Province of Ontario.

\end{document}